\documentclass[prl,aps,preprint]{revtex4}

\usepackage{amssymb}     
\usepackage{graphicx}    
\usepackage{ifthen}
\usepackage[english]{babel}
\usepackage{amsmath}

\begin{document} 

\bibliographystyle{unsrt}
\title{Orbital quantization in the high magnetic field state 
of a charge-density-wave system} 
\author{D. Andres$^1$, M. V. Kartsovnik$^1$, P. D. Grigoriev$^{2}$,  
W. Biberacher$^1$ and H. M\"{u}ller$^3$} 
\affiliation{$^{1}$ Walther-Mei\ss ner-Institut, Bayerische Akademie der 
Wissenschaften, Walther-Mei\ss ner-Str. 8, D-85748 Garching, Germany\\ 
$^{2}$ L. D. Landau Institute for Theoretical Physics, 142432 Chernogolovka, 
Russia  \\ 
$^3$ European Synchrotron Radiation Facility, F-38043 Grenoble, France
 } 

\date{\today } 

\begin{abstract} 
 A superposition of the Pauli and orbital coupling of a high magnetic 
field to charge carriers in a charge-density-wave (CDW) system is 
proposed to give rise to transitions between subphases with quantized 
values of the CDW wavevector. By contrast to the purely orbital 
field-induced density-wave effects which require a strongly imperfect 
nesting of the Fermi surface, the new transitions can occur even if 
the Fermi surface is well nested  at zero field. We suggest that such 
transitions are observed in the organic metal 
$\alpha$-(BEDT-TTF)$_2$KHg(SCN)$_4$ under a strongly tilted magnetic 
field. 
\end{abstract} 
\maketitle
\mbox{} 

A quantization of the orbital motion of charge carriers in 
quasi-one-dimensional (Q1D) organic conductors (TMTSF)$_2$X under high 
magnetic fields leads to a remarkable quantum phenomenon  of field-induced 
spin-density waves (FISDW) (see \cite{Ishi,cha96} for a review). 
In these salts, the conducting band can be described near the Fermi level by 
the simplified dispersion law:
\begin{equation} 
\epsilon ({\mathbf k})=\hbar v_{F}\left( \left| k_{x}\right| -k_{F}\right) 
-2t_{y}\cos(k_y a_y)-2t_{y}^{\prime }\cos (2k_y a_y) 
-2t_{z}\cos(k_z a_z),   \label{ESQ1D} 
\end{equation} 
where $\frac{t_y}{E_F} \sim \frac{t_y^{\prime}}{t_y} \simeq 0.03$; 
$\frac{t_z}{E_F} \sim 10^{-3}$; $E_F$, $v_F$, and $k_F$ are, respectively, 
the Fermi energy, Fermi velocity and Fermi wavenumber in the chain 
direction $x$; 
$a_y$ and $a_z$ are the interchain distances. Correspondingly, the Fermi 
surface (FS) is 
represented by two flat, only slightly warped sheets extended in $k_yk_z$ 
plane. At small $t_y^{\prime}$ such a FS is subject to a nesting instability 
resulting in an insulating charge-density wave (CDW) or spin-density wave 
(SDW) groundstate \cite{Ishi}. 
In the (TMTSF)$_2$X salts the groundstate is a SDW. One can increase the 
``antinesting'' term $t_y^{\prime}$, e.g. 
by applying pressure, thereby weakening the nesting instability. 
At $t_y^{\prime}>t_y^{\prime \ast} \simeq k_BT_c(0)$ [where 
$T_c(0)$ is the density-wave transition temperature at $t_y^{\prime}=0$] the 
SDW is completely suppressed \cite{dan96,yam82,mon88}. However, a magnetic field ${\mathbf B}$ applied 
along the 
$z$-axis restricts electron orbits to the chain direction, i.e. enhances 
their one-dimensionality, and restores the SDW state 
\cite{cha96,mon88,kwa86,gor84}. 
Since the FS nesting is 
strongly imperfect, the new SDW state is semimetallic rather than insulating. 
Under a high magnetic field the semimetallic spectrum is subdivided into 
Landau subbands and the system gains energy when the Fermi level is situated 
between adjacent subbands. This can be achieved by an appropriate choice of 
the nesting vector ${\mathbf Q}$ which determines the size of the 
semimetallic pockets.  
As a result, ${\mathbf Q}$ becomes field-dependent, deviating from its 
ideal value 
${\mathbf Q}_0=(2k_F,\pi/a_y,\pi/a_z)$, and at low enough temperatures the 
system undergoes a series of first order phase transitions to FISDW subphases 
with quantized values of the $Q_x$ component 
\cite{Ishi,cha96,gor84,mon85,leb02,sen03}: 
\begin{equation}
Q_{xN} = 2k_F + NG, \label{OQ}
\end{equation}
where $N$ is an integer and $G=2ea_yB_z/\hbar c$. 

Similarly, the orbital effect has been predicted to produce 
field-induced charge-density wave (FICDW) subphases \cite{zanc,leb03}. 
The theoretical predictions are supported by recent 
pressure-dependent magnetotransport studies of the organic 
conductor $\alpha $-(BEDT-TTF)$_2$KHg(SCN)$_4$ \cite{and01} which is believed 
to be in a CDW state with a nested Q1D part of its FS below $\simeq 8.5$ K 
\cite{mck97,bis98,chr00,fou03}. 
However, in comparison with the SDW case, the situation in CDW systems is 
more complicated due to the presence of an additional, Pauli effect. The CDW 
formation is based on the pairing interaction 
between the electrons and holes having the same spin and separated in 
$k$-space by a unique wavevector (nesting vector) $\mathbf{Q}$. 
A magnetic field lifts the degeneracy between two subbands with the 
parallel and antiparallel spins, respectively, so that the zero-field 
wavevector does not provide optimal nesting conditions 
for either of the subbands. This leads to a decrease of the critical 
temperature $T_c(B)$ \cite{die73}. 
It is mainly the Pauli effect which determines the shape of the $B-T$ phase 
diagram of $\alpha $-(BEDT-TTF)$_2$KHg(SCN)$_4$ at ambient pressure, in 
the field perpendicular to the highly conducting $ac$-plane 
\cite{bis98,chr00}. However, the orbital effect was also suggested to play a 
significant role under pressure \cite{and01} or in a magnetic field strongly 
tilted towards the $ac$-plane \cite{qua00}. 

In this Letter, we propose that 
the superposition of Pauli and orbital effects of a high magnetic 
field on a CDW can give rise to a series of transitions between subphases 
with quantized values of the nesting vector. 
By contrast to the above mentioned FISDW and FICDW effects, these new  
transitions do not require a suppressed density wave state 
at zero magnetic field ($t_y^{\prime}>t_y^{\prime 
\ast}$). We analyze the complex behavior of $\alpha$-(BEDT-TTF)$_2$KHg(SCN)$_4$ 
in tilted magnetic fields and argue that the proposed phenomenon is 
realized in this compound.

The field-induced transitions in single crystals of 
$\alpha $-(BEDT-TTF)$_2$KHg(SCN)$_4$ were studied by simultaneous 
measurements of magnetic torque and resistivity versus magnetic field as 
described elsewhere \cite{wei99}. The resistivity was measured in the 
$b^*$ direction ($b^* \perp ac$).
The sample could be turned with respect to the magnetic field and its 
orientation was defined by the tilt angle $\theta$ between 
the $b^*$-axis and the field direction. The absolute value of $\theta$ was 
determined to the accuracy of $\leq 0.1^{\circ}$. This was assured by measuring 
the frequency of the de Haas-van Alphen (dHvA) oscillations originating from 
the second, quasi-two-dimensional, conducting band which precisely follows 
the $1/\cos \theta $-law.

We start with a qualitative consideration of the field dependence of the 
$x$ component of the nesting vector for a CDW system described by the 
dispersion relation (\ref{ESQ1D}) with a negligibly small $t_z$ and 
$0<t_y^{\prime}<t_y^{\prime \ast}$.
Treating each spin subband independently, 
one can express the optimal nesting conditions as
$Q_{{\text{opt,}}x}^{\uparrow (\downarrow )}(B)=Q_{0x} \pm 2\mu_B B/\hbar 
v_F$, where the sign $+(-)$ stands for the spins parallel (antiparallel) 
to the applied field (see dashed lines in Fig.1). 
Nevertheless, the system as a whole maintains the constant 
nesting vector ${\mathbf Q}_0$ and both subbands will remain fully gapped
 up to the critical field $B_k \sim \Delta_{\text{CDW}}/2 \mu _B$, 
 where $\Delta_{\text{CDW}}$ is the CDW energy gap \cite{zanc}. 
Above $B_k$, ${\mathbf Q}_0$ is no more a good nesting vector as it leads to 
ungapped states in both subbands. 
Then, at low enough temperatures, $T<T^{\ast}\simeq 0.6T_c(0)$, 
a CDW$_x$ state with the field-dependent nesting vector, 
$Q_x(B)=Q_{0x}+q^{\text{Pauli}}_x(B)$, is formed \cite{zanc,comment1}. 
The additional term $q^{\text{Pauli}}_x(B)$ asymptotically approaching the 
value $2\mu _B B/\hbar v_F$ is schematically shown in Fig. 1 by the thin 
solid line. 
With this term the nesting conditions are obviously improved for one of the 
spin subbands (say, the spin-up subband) at the cost of an additional 
unnesting of the other (spin-down). 

So far the Pauli effect was considered as a mechanism dominating the            
evolution of the CDW in a magnetic field.                                       
If the antinesting term $t_y^{\prime}$ is sufficiently lower than the           
critical value $t_y^{\prime \ast}$, the orbital effect below $B_k$ 
only leads to a small increase of the CDW gap with increasing field.
However, for $B>B_k$ one should take into account that the carriers of the spin-down 
subband become partially ungapped. If the field is not too high, this subband 
forms small, closed in $k_xk_y$ plane pockets \cite{comment2}. 
The situation resembles the ``conventional'' FISDW or 
FICDW cases with $t_y^{\prime} \gtrsim t_y^{\prime \ast}$; only the 
unnesting role is taken by the Pauli effect. Therefore, one can expect 
the orbital quantization condition in the form of Eq.(\ref{OQ}) to be set on the 
nesting vector. However, the quantized levels should be now counted from 
$Q_{{\text{opt,}}x}^{\downarrow }(B)$:
\begin{equation} 
Q_{x,N}^{\text{orbit}} = Q_{0x} - 2\mu_BB/\hbar v_F + NG. \label{OQ1} 
\end{equation} 
The corresponding values $q_{x,N}^{\text{orbit}}= Q_{x,N}^{\text{orbit}} 
- Q_{0x}$ are schematically shown by dotted lines in Fig. 1. 

As a result, the most favorable values of the nesting vector above $B_k$ are 
determined by intersections of the continuous curve $q^{{\text{Pauli}}}_x(B)$ 
with the straight lines $q^{{\text{orbit}}}_{xN}$, i.e. by the superposition 
of the Pauli and quantum orbital effects. Thus, with changing the field we obtain a series of first order transitions between CDW subphases 
characterized by different quantized values of the nesting vector as 
schematically shown by thick lines in Fig. 1. Note that our model implies an 
increase of the quantum number $N$ with increasing $B$, in contrast to what 
is usually observed in known orbital quantization phenomena. 

The presented qualitative model will now be compared with the experimental 
results on $\alpha $-(BEDT-TTF)$_2$KHg(SCN)$_4$. 
Fig. 2 shows magnetic torque and interlayer resistance 
recorded at increasing (solid lines) and decreasing (dashed lines) field 
sweeps, at $\theta = 81.4^{\circ}$. In agreement with previous reports 
\cite{chr00,qua00,chr96}, 
below 2 K both quantities consistently display irregular oscillations with 
a significant hysteresis. Both the amplitude of the oscillations and strength 
of the hysteresis rapidly increase with decreasing temperature. The so-called 
kink transition at $B_k\approx 12.7$ T is supposed to be the transition 
between the CDW$_0$ groundstate and CDW$_x$ \cite{and01,mck97,bis98,chr00}. This transition 
is considerably shifted from its zero-angle value $B_k(\theta=0) \approx 
24$ T (see inset in Fig. 2 for comparison). In addition, there are 
two features at higher fields reminiscent of first order phase transitions. 

In order to clarify the origin of the observed structure, we have studied 
in detail how it develops with changing $\theta$. In the following we 
analyze the torque data divided by $\sin(2\theta)$ in order to enable a 
direct comparison between the curves recorded at different $\theta$'s. 
The features in the magnetoresistance 
are found to correlate with those in the torque (see 
e.g. Fig. 2) although at $\theta$ approaching $90 ^{\circ}$ they become more 
smeared and difficult to interpret in terms of phase transitions. 

Fig. 3 shows the field-dependent torque at different angles, 
$T=0.4$ K (the curves are vertically shifted with respect to each other for 
clarity). The data represented by solid lines 
have been taken at increasing $B$; additionally, 
down sweeps are shown for $81^{\circ} < \theta < 84^{\circ}$ (dashed 
lines) to illustrate the hysteresis. 
The curves corresponding to $\theta < 81^{\circ}$ have been obtained by 
filtering out the dHvA oscillations from the measured signal. 
An example of the total signal, including the oscillations, at 
$\theta = 67.1^{\circ}$ is given by the grey dotted line in Fig. 3a. 

At relatively low angles, one can see a sharp kink transition which takes 
place at $\approx 22.5$ T at 53$^{\circ}$. It gradually shifts down, 
saturating at $\simeq 10$ T as $\theta$ approaches 90$^{\circ}$. As $B_k$ 
decreases below 20 T, the transition becomes less pronounced although can 
still be detected up to $\theta \approx 85^{\circ}$. Below $B_k$, a weak 
additional feature is observed up to 78$^{\circ}$ (dotted line in Fig. 3a). 
While its origin is not 
clear at present, it may be related to the hysteretic changes of the 
magnetoresistance \cite{har99,pro00} and magnetothermopower \cite{cho02} 
observed in the same field range earlier.

Starting from 65$^{\circ}$, new features emerge at the 
high-field side, moving to lower fields with further increasing $\theta$.
 Like the kink transition, 
they are rather sharp at high fields and gradually fade below 20 T. The 
structure changes with increasing rate at $\theta \rightarrow 90^{\circ}$. 
At the same time the features become less pronounced and could not be 
resolved above $88^{\circ}$. While the observed behavior is strongly 
suggestive of multiple phase transitions, at the moment it is difficult to 
determine the exact location of each transition. However, taking into account 
the character of changes in the magnetization and the hysteresis behavior, it 
is reasonable, by analogy with $B_k$, to identify the 
transition points with local maxima in the derivative $(\partial M/\partial 
B)_{\theta }$. 

Such points are plotted in Fig. 4 in the form of a $B-\theta $ phase 
diagram. In the inset the same data are replotted against $1/\cos \theta$ in 
order to show the region near $90^{\circ}$ in more detail. The crosses 
represent $B_k(\theta )$, the empty circles correspond to the weak feature 
below $B_k$, and the solid circles delineate the boundaries between new 
subphases within the high-field state. There is a clear tendency for all the 
transitions to shift to lower fields with increasing $\theta$. Further, the 
increasingly strong angular dependence in the vicinity of 
$\theta =90^{\circ}$ and the absence of the features at the exactly parallel 
orientation suggest that the orbital effect determined by the perpendicular 
component of magnetic field $B_{z}=B\cos \theta$ plays a crucial role. On the 
other hand, the CDW state is obviously subject to the strong Pauli effect at 
$B>B_k$ \cite{zanc}. Therefore it is very likely that the observed 
transitions originate from an interplay between the Pauli and orbital effects 
on the high field CDW state.

From Fig. 1 it is clear that the proposed multiple transitions can only take 
place if more than one quantum level intersect the $q_x^{\text{Pauli}}(B)$ 
curve, i.e. when $3G=6ea_yB_z/\hbar c < 4\mu _BB/\hbar v_F$. This condition 
is obviously not satisfied at low $\theta$ where only one, kink transition is 
observed. With increasing $\theta$, the separation between the levels, $G$, 
proportional to the perpendicular field component $B_z$ 
decreases whereas the isotropic Pauli effect remains unchanged. Starting from 
a threshold angle $\theta ^{\ast}$, the above condition becomes fulfilled and 
a new transition emerges at a high field, shifting towards $B_k$ at a further 
increase of $\theta$. In our experiment the first transition above $B_k$ is 
found at $\theta \approx 65^{\circ}$. This allows us to estimate the upper 
limit for the Fermi velocity in the 1D direction: $v_F \lesssim 
2\mu _Bc/3ea_y\cos 65^{\circ} \approx 9\times 10^6$~cm/sec, which is in good 
agreement with the recently reported value $v_F = 6.5 \times 10^6$~cm/sec 
 \cite{kov02}. Thus, the proposed model explains not only the 
occurrence of multiple first-order transitions above $B_k$, but also
the threshold angle above which they start to appear as well as the general 
tendency of them to shift towards $B_k$ with further tilting the field. 

Obviously, the sharpness of the transitions between the quantum orbital 
levels should strongly depend on temperature and on the strength of the 
perpendicular field component. This is consistent with the observed damping 
of the structure with increasing temperature (Fig. 2) or at $\theta$ very 
close to $90^{\circ}$ (Fig. 3b). 

The experimental phase diagram (Fig. 4) looks more complicated than it would 
be expected within the present qualitative model: some phase lines appear to 
be split, giving rise to additional subphases. This disagreement is, however, 
not surprising. On the one hand, it is not excluded that the additional 
phase, CDW$_y$ ($q_x=0; q_y \neq 0$), which was not taken into account here, emerges at a certain 
angular and field range \cite{qua00,zanc}. On the other hand, the orbital quantization effects 
 are known to strongly depend on the exact FS geometry, i.e on details of 
the antinesting term, in the case of FISDW \cite{zan96b}. According to the 
band structure calculations \cite{band}, the antinesting term in the present 
compound must be  significantly more complex than suggested by the model 
spectrum (\ref{ESQ1D}). 
Therefore it is expected to also lead to considerable 
complications in high-field subphase structure.
There are other open questions, such as the transition-like feature below 
$B_k$ and the considerable (although at a slower rate) decrease of $B_k$ 
with increasing $\theta$. A development of a quantitative theory will 
certainly help to understand the entire phase diagram of the present compound 
and to get a deeper insight into the interplay of the Pauli and orbital 
effects on CDW. 

The experimental data presented in this paper have been obtained at the 
Grenoble High Magnetic Field Laboratory. The work was supported by 
HPP Programme of EU, contracts HPRI-1999-CT-00030 and 
HPRI-CT-1999-40013, INTAS grant 01-0791, and RFBR 03-02-16121.

\newpage 

\begin{center} 
Figure captions 
\end{center}

Fig. 1. Schematic illustration of the field dependence of the CDW wavevector 
(thick line) originating from the superposition of the Pauli effect (thin 
solid line) \cite{zanc} and orbital quantization. The quantized orbital 
levels (dotted lines) are counted from $q_{{\text{opt,}}x}^{\downarrow }(B)$ 
corresponding to the optimal nesting vector for the spin-down subband which 
is partially ungapped above $B_k$. 
 
Fig. 2. (a) Up (solid lines) and down (dashed lines) field sweeps of the 
magnetic torque of $\alpha $-(BEDT-TTF)$_2K$Hg(SCN)$_4$ at $\theta 
=81.4^{\circ}$, at $T=0.4$ and 1.3 K. (b) The same for the interlayer 
resistance. Vertical line indicates the kink transition field $B_k$. 
Inset: magnetoresistance at a low $\theta$, $T=0.4$ K. 

Fig. 3. Evolution of the high-field structure in the steady torque with 
changing $\theta$. Solid lines: up sweeps of the field; dashed lines 
(at $81^{\circ} <\theta <84^{\circ}$): down sweeps. For $\theta < 81^{\circ}$ 
the dHvA signal is subtracted; an example of the total signal is given by 
the grey dotted line for $\theta = 67.1^{\circ}$. 

Fig. 4. $B-\theta$ phase diagram of $\alpha $-(BEDT-TTF)$_2K$Hg(SCN)$_4$. 
Crosses: kink transition $B_k$; open circles: weak feature below $B_k$; 
solid circles: multiple transitions inside the high-field state. 
Lines are guides to the eye. 


\newpage
\begin{thebibliography}{10}                                                          
\expandafter\ifx\csname 
bibnamefont\endcsname\relax 

\def\bibnamefont#1{#1}\fi 
\expandafter\ifx\csname 
bibfnamefont\endcsname\relax 

\def\bibfnamefont#1{#1}\fi 
\expandafter\ifx\csname 
url\endcsname\relax 
  \def\url#1{\texttt{#1}}\fi 
\expandafter\ifx\csname urlprefix\endcsname\relax\def\urlprefix{URL }\fi 
\expandafter\ifx\csname bibinfo\endcsname\relax \def\bibinfo#1#2{#2}\fi 
\expandafter\ifx\csname eprint\endcsname\relax \def\eprint#1{#1}\fi 


\bibitem{Ishi} 
\bibinfo{author}{\bibfnamefont{T.}~\bibnamefont{Ishiguro}}, 
\bibinfo{author}{\bibfnamefont{K.}~\bibnamefont{Yamaji}}, 
\bibnamefont{and}  
\bibinfo{author}{\bibnamefont{G.~Saito}}, 
\emph{\bibinfo{title}{Organic Superconductors, 2$^{nd}$ edition}} 
(\bibinfo{publisher}{Springer-Verlag Berlin Heidelberg}, 
\bibinfo{year}{1998}). 

\bibitem{cha96} 
\bibinfo{author}{\bibfnamefont{P.~M.}~\bibnamefont{Chaikin}}, 
\bibinfo{journal}{J. Phys. I France} 
\textbf{\bibinfo{volume}{6}}, 
\bibinfo{pages}{1875} 
(\bibinfo{year}{1996}). 

\bibitem{dan96} G.~M.~Danner, P.~M.~Chaikin, and S.~T.~Hannahs, Phys. Rev. B 
{\bf 53}, 2727 (1996). 

\bibitem{yam82} K.~Yamaji, J. Phys. Soc. Jpn. {\bf 51}, 2787 (1982); 
 Y.~Hasegawa and H.~Fukuyama, J. Phys. Soc. Jpn. {\bf 55}, 3978 (1986). 
 
\bibitem{mon88} G.~Montambaux, Phys. Rev. B {\bf 38}, 4788 (1988). 

\bibitem{kwa86} J.~F.~Kwak et al., 
Phys. Rev. Lett. {\bf 56}, 972 (1986). 

\bibitem{gor84} L.~P.~Gor'kov and A.~G.~Lebed, J. Phys. (Paris) Lett. 
{\bf 45}, L433 (1984). 

\bibitem{mon85} G.~Montambaux, M.~H\'{e}ritier, and P.~Lederer, Phys. Rev. 
Lett. {\bf 55}, 2078 (1985). 

\bibitem{leb02} 
\bibinfo{author}{\bibfnamefont{A.~G.} 
\bibnamefont{Lebed}}, 
\bibinfo{journal}{Phys. Rev. Lett.} 
\textbf{\bibinfo{volume}{88}}, 
\bibinfo{pages}{177001} (\bibinfo{year}{2002}); 
A.~G.~Lebed, Pis'ma Zh. Teor. Eksp. Fiz. {\bf 72}, 205 (2000) [JETP Lett. 
{\bf 72}, 141 (2000)]. 

\bibitem{sen03} K.~Sengupta and N.~Dupuis, cond-mat/0302142, unpublished. 

\bibitem{zanc} 
\bibinfo{author}{\bibfnamefont{D.}~\bibnamefont{Zanchi}}, 
\bibinfo{author}{\bibfnamefont{A.}~\bibnamefont{Bjeli\v{s}}}, 
\bibnamefont{and} 
\bibinfo{author}{\bibfnamefont{G.}~\bibnamefont{Montambaux}}, 
\bibinfo{journal}{Phys. Rev. B} 
\textbf{\bibinfo{volume}{53}}, 
\bibinfo{pages}{1240} 
(\bibinfo{year}{1996}). 

\bibitem{leb03} A.~G.~Lebed, Synth. Metals in press. 

\bibitem{and01} 
\bibinfo{author}{\bibfnamefont{D.}~\bibnamefont{Andres et al.}}, 
\bibinfo{journal}{Phys. Rev. B} 
\textbf{\bibinfo{volume}{64}}, 
\bibinfo{pages}{161104(R)} 
(\bibinfo{year}{2001}). 

\bibitem{mck97} R.~H.~McKenzie, cond-mat/9706235, unpublished. 

\bibitem{bis98} 
\bibinfo{author}{\bibfnamefont{N.}~\bibnamefont{Biskup}}, 
\bibinfo{author}{\bibfnamefont{J.~A.~A.~J.}~\bibnamefont{Perenboom}}, 
\bibinfo{author}{\bibfnamefont{J.~S.}~\bibnamefont{Qualls}}, 
\bibnamefont{and} 
\bibinfo{author}{\bibfnamefont{J.~S.}~\bibnamefont{Brooks}}, 
\bibinfo{journal}{Solid State Commun.} 
\textbf{\bibinfo{volume}{107}}, \bibinfo{pages}{503} 
(\bibinfo{year}{1998}).   


\bibitem{chr00} 
\bibinfo{author}{\bibfnamefont{P.}~\bibnamefont{Christ et al.}}, 
\bibinfo{journal}{JETP Lett.} 
\textbf{\bibinfo{volume}{71}}, 
\bibinfo{pages}{303} 
(\bibinfo{year}{2000}). 

\bibitem{fou03} 
\bibinfo{author}{\bibfnamefont{P.}~\bibnamefont{Foury-Leylekian}}, 
\bibinfo{author}{\bibfnamefont{S.}~\bibnamefont{Ravy}},
\bibinfo{author}{\bibfnamefont{J.~P.}~\bibnamefont{Pouget}}, 
\bibnamefont{and} 
\bibinfo{author}{\bibfnamefont{H.}~\bibnamefont{M\"uller}}, 
\bibinfo{journal}{Synth. Met., to be published}. 

\bibitem{die73} 
\bibinfo{author}{\bibfnamefont{W.}~\bibnamefont{Dieterich}} 
\bibnamefont{and} 
\bibinfo{author}{\bibfnamefont{P.}~\bibnamefont{Fulde}}, 
\bibinfo{journal}{Z. Physik} 
\textbf{\bibinfo{volume}{265}}, \bibinfo{pages}{239} 
(\bibinfo{year}{1973}). 
                                                                                     \bibitem{qua00} 
\bibinfo{author}{\bibfnamefont{J.~S.}~\bibnamefont{Qualls et al.}}, 
\bibinfo{journal}{Phys. Rev. B} 
\textbf{\bibinfo{volume}{62}}, 
\bibinfo{pages}{10008} 
(\bibinfo{year}{2000}). 

\bibitem{wei99} H.~Weiss et al., 
Phys. Rev. B {\bf 60}, R16259 (1999). 

\bibitem{comment1} We do not consider here an 
additional, CDW$_{y}$ state with a field-dependent $y$ component of 
the nesting vector which was suggested to exist in a narrow field range 
for certain values of electron coupling constants \cite{zanc}. This, however, 
does not affect the physical nature of the described phenomenon. 

\bibitem{comment2}                                                                      Based on geometrical considerations, one can roughly          
estimate the upper limit for the field, at which closed pockets are formed,   
from the condition: $2\mu_BB + \hbar q_x^{\text{Pauli}}v_F - 
\Delta_{\text{CDW}}< 2t_y^{\prime}$. 

\bibitem{chr96} P.~Christ et al., 
Synth. Metals {\bf 70}, 823 (1996). 

\bibitem{har99} N.~Harrison et al., 
Phys. Rev. B {\bf 55}, R16005 (1997).
                                                                                     
\bibitem{pro00} 
\bibinfo{author}{\bibfnamefont{C.}~\bibnamefont{Proust et al.}}, 
\bibinfo{journal}{Phys. Rev. B} 
\textbf{\bibinfo{volume}{62}}, 
\bibinfo{pages}{2388} 
(\bibinfo{year}{2000}). 

\bibitem{cho02} 
\bibinfo{author}{\bibfnamefont{E.~S.} 
\bibnamefont{Choi}}, 
\bibinfo{author}{\bibfnamefont{J.~S.} 
\bibnamefont{Brooks}}, 
\bibnamefont{and} 
\bibinfo{author}{\bibfnamefont{J.~S.} 
\bibnamefont{Qualls}}, 
\bibinfo{journal}{Phys. Rev. B} 
\textbf{\bibinfo{volume}{65}}, 
\bibinfo{pages}{205119} 
(\bibinfo{year}{2002}). 

\bibitem{kov02} 
\bibinfo{author}{\bibfnamefont{A.~E.} 
\bibnamefont{Kovalev}}, 
\bibinfo{author}{\bibfnamefont{S.}~\bibnamefont{Hill}}, 
\bibnamefont{and} 
\bibinfo{author}{\bibfnamefont{J.~S.} 
\bibnamefont{Qualls}}, 
\bibinfo{journal}{Phys. Rev. B} 
\textbf{\bibinfo{volume}{66}}, 
\bibinfo{pages}{134513} 
(\bibinfo{year}{2002}). 

\bibitem{zan96b} 
\bibinfo{author}{\bibfnamefont{D.}~\bibnamefont{Zanchi}} 
\bibnamefont{and} 
\bibinfo{author}{\bibfnamefont{G.}~\bibnamefont{Montambaux}}, 
\bibinfo{journal}{Phys. Rev. Lett.} 
\textbf{\bibinfo{volume}{77}}, 
\bibinfo{pages}{366} 
(\bibinfo{year}{1996}). 

\bibitem{band} R.~Rousseau et al., 
J. Phys. I France {\bf 6}, 1527 (1996); M.~Gusm\~{a}o and T.~Ziman, 
Phys. Rev. B {\bf 54}, 16663 (1996). 

\end{thebibliography}
\end{document}